\documentclass[twocolumn]{autart}    

\usepackage{amsmath,amssymb}
\usepackage{mathrsfs}
\usepackage{graphicx}          

\begin{document}

\begin{frontmatter}

\title{Scheduler-Pointed False Data Injection Attack for Event-Based Remote State Estimation\thanksref{footnoteinfo}} 

\thanks[footnoteinfo]{This work was supported by the
	National Natural Science Foundation of China under Grant 61773357. The material in this paper was not presented at any conference. Corresponding author Junlin Xiong.}

\author[Hefei]{Qiulin Xu}\ead{xuqiulin@mail.ustc.edu.cn},    
\author[Hefei]{Junlin Xiong}\ead{junlin.xiong@gmail.com},               

\address[Hefei]{Department of Automation, University of Science and Technology of China, Hefei 230026, China}  


\begin{abstract}                          
In this paper, an attack problem is investigated for event-based remote
state estimation in cyber-physical systems. Our objective is to degrade the
effect of the event-based scheduler while bypassing a $\chi^2$ false data
detector. A two-channel scheduler-pointed false data injection attack strategy is proposed by modifying
the numerical characteristics of innovation signals. The attack strategy is proved to
be always existent, and an algorithm is provided to find it.
Under the proposed attack strategy, the scheduler becomes almost invalid and
the performance of the remote estimator is degraded.
Numerical simulations are used to illustrate our theoretical results.
\end{abstract}

\end{frontmatter}

\section{Introduction}
Recently, the security problem of cyber-physical systems
(CPSs) has attracted great attention. Successful attacks lead to a variety of
serious consequences, such as negative impact on economy, national
security and even human lives\cite{poovendran2011special}. The existing attack strategies can be
roughly classified into two groups: denial-of-service (DoS) attacks and
deception attacks. DoS attacks aim to violate data availability through
obstructing the transmission of information flows
\cite{zhang2015optimal}. Deception attacks affect the integrity
of the transmitted data  while remaining stealthy to the anomaly
detection \cite{liu2011false}. Compared with DoS
attacks, deception attacks are more energy-saving and stealthy, and fit the aim of
malicious agents\cite{teixeira2015secure}.




Deception attacks have recently received much attention, and can be divided
into two types. The first one is called reply attack where sensor data are
recorded and replayed without system knowledge. The feasibility problem was
studied for control systems equipped with false-data
detectors\cite{mo2009secure} and a countermeasure was proposed in
\cite{mo2013detecting} to detect such an attack. The second is called
false-data injection (FDI) attack where system knowledge is required. The idea
of FDI attack is initially proposed in \cite{liu2011false} for power network
systems. In \cite{guo2016optimal}, an optimal linear FDI attack strategy was
designed for remote state estimation problems. A two-channel attack strategy
was proposed to handle more general situations in~\cite{pang2016two}. The
authors of \cite{ni2019performance} considered a stealthy attack, which could
drive the system state to some other target states instead of driving to
infinity in the most of existing results\cite{hu2018state}.

In CPSs, both packet dropout and event-based schedulers are common for the communication networks. As a result, the sensor-to-estimator communication rate is not $100\%$ and the above results\cite{guo2016optimal,pang2016two,ni2019performance,hu2018state} become no longer
applicable. For example, the strategies in
\cite{ni2019performance,hu2018state} are based on idealistic communication
channels. Naturally, an interesting topic is to consider FDI attack problem in
more realistic situations.

In this paper, we propose a new type of FDI attack strategy for remote state
estimation problem with an event-based data scheduler. In practice,
event-based schedulers are widely used to reduce communication rate for saving
energy\cite{mo2013detecting} or sparing transmission
bandwidth\cite{shi2014event}. For example, a event-based scheduler for remote
state estimation was proposed in \cite{wu2012event}, where the
sensor schedule was determined by innovation of
Kalman filters. Our work aims at injecting false data to degrade the
event-based scheduler without being detected by a false data detector at the remote state estimator. We propose a two-channel scheduler-pointed FDI attack strategy. In the forward channel, the innovation sequence
is attacked through scaling up the mean to trigger the scheduler while squeezing
the variance to bypass the detector. In the feedback channel, the feedback
data is modified to guarantee the effect of the forward attack. The existence of our attack
strategy is proved by characteristics of Gaussian distribution, and the feasible set of all attack
parameters can be found analytically. We also analyzed the performance
evolution of the remote estimator under this type of attacks. Our result
shows that a successful attack strategy can always be
designed such that the event-based scheduler becomes almost invalidated and
the estimation performance is degraded simultaneously. Finally, a
numerical example is presented to illustrate the efficiency of the attack
strategy.

\textit{Notation:} $\mathbb{N}$ and $\mathbb{R}$ denote the sets of natural numbers and real numbers, respectively. $\mathbb{R}^{n}$ is the $n$-dimensional Euclidean space. $\mathbb{S}_{+}^{n}$ is the set of $n \times n$ positive semi-definite matrices. When $X \in \mathbb{S}_{+}^{n}$, we simply write $X \geq 0$; Similarly, $X \geq Y$ means $X-Y \geq 0$. $\mathcal{N}(\mu, \Sigma)$ denotes Gaussian distribution with mean $\mu$ and covariance matrix $\Sigma$. $\mathbf{E}[\cdot]$ denotes the mathematical expectation and $\mathrm{Pr}(\cdot)$ denotes the probability of a random event. $\mathrm{Tr}\{\cdot\}$ denotes the trace of a matrix and the superscript ``${\top}$'' stands for transposition.  $I_{n}$ is the $n \times n$-dimensional identity matrix. The term $x x^{\top}$ for vector $x$ is abbreviated as $x (*)^{\top}$.

\section{Preliminaries and Problem Formulation}
This section firstly describes the model of an event-based remote state estimator as shown in Fig.~\ref{fig0}. A two-channel scheduler-pointed FDI attack strategy is then formulated.

\subsection{Process Model}
\begin{figure}
\begin{center}
\includegraphics[width=3.5in]{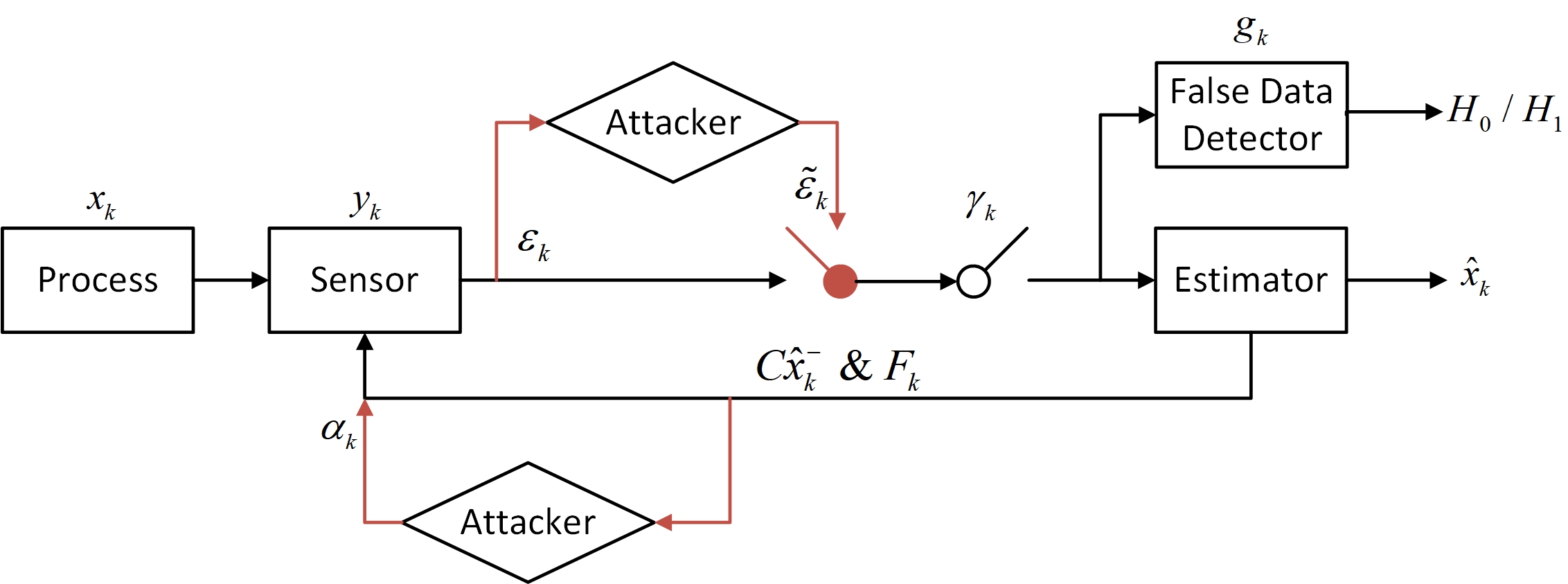}    
\caption{System block diagram.}  
\label{fig0}                                 
\end{center}                                 
\end{figure}

The physical plant in Fig. \ref{fig0} is described by a discrete time linear time-invariant process
\begin{align}
x_{k+1}&=Ax_k+w_k  \label{1} \\
y_{k}&=Cx_k+v_k  \label{2}
\end{align}
where $k \in \mathbb{N}$ is the time index, $x_{k} \in \mathbb{R}^{n}$ is the vector of system states, $y_{k} \in \mathbb{R}^{m}$ is the vector of sensor measurements, $w_{k} \in \mathbb{R}^{n}$ and $v_{k} \in \mathbb{R}^{m}$ are zero-mean i.i.d. Gaussian noises with covariances $Q \geq 0$ and $R>0,$ respectively. The initial state $x_{0}$ is zero-mean Gaussian with covariance $\Xi_{0} \geq 0 $, and is independent of $w_{k}$ and $v_{k}$ for all $k \geq 0 .$ The pair $(A, C)$ is detectable and $(A, \sqrt{Q})$ is stabilizable.

\subsection{Event-Based Remote Estimator}
Based on sensor data and estimator feedback data, an event-based scheduler and the corresponding approximate minimum mean-squared error (MMSE) estimator are shown as follows\cite{wu2012event}:
\begin{itemize}
	\item[1)]\textit{Time update}:
	\begin{equation}
	\left\{\begin{aligned}
	\hat{x}_{k}^{-}&=A \hat{x}_{k-1} \\
	P_{k}^{-}&=h\left(P_{k-1}\right)
	\end{aligned}\right. \label{3}
	\end{equation}
	where
	\begin{equation}
	h(X) \triangleq A X A^{\top}+Q .
	\end{equation}
	The estimates $\hat{x}_{k}^{-}$ and $\hat{x}_{k}$ here are \textit{a priori} and \textit{a posteriori} MMSE estimate, respectively; and the matrices $P_{k}^{-}$ and $P_{k}$ are the corresponding error covariances.
	\item[2)]\textit{Measurement update}:
	\begin{equation}
	\left\{\begin{aligned}
	\hat{x}_{k}&=\hat{x}_{k}^{-}+\gamma_{k} K_{k} z_{k} \\
	K_{k} &= P_{k}^{-} C^{\top}\left[C P_{k}^{-} C^{\top}+R\right]^{-1} \\
	P_{k}&=\gamma_{k} \tilde{q}\left(P_{k}^{-}\right)+\left(1-\gamma_{k}\right) \tilde{q}_{\kappa(\beta)}\left(P_{k}^{-}\right)
	\end{aligned}\right. \label{5}
	\end{equation}
	where $z_{k}$ is the measurement innovation, $\gamma_{k}$ is the binary decision variable indicating whether the event-based scheduler shall be closed ($\gamma_{k} = 1$) or open ($\gamma_{k} = 0$), $K_{k}$ is the standard Kalman gain, and
	\begin{align}
	\tilde{q}_{\lambda}(X) &\triangleq X-\lambda X C^{\top}\left[C X C^{\top}+R\right]^{-1} C X \\
	\kappa(\beta)&=\frac{2}{\sqrt{2 \pi}} \beta e^{-\frac{\beta^{2}}{2}}[1-2 Q(\beta)]^{-1}
	\end{align}
	if $\lambda = 1$, $\tilde{q}_{\lambda}$ will be written as $\tilde{q}$ for brevity. $Q(\cdot)$ is the standard $Q$-function defined by
	\begin{equation}
	Q(\beta) \triangleq \int_{\beta}^{+\infty} \frac{1}{\sqrt{2 \pi}} e^{-\frac{x^{2}}{2}} \mathrm{d} x .
	\end{equation}
	\item[3)]\textit{Mahalanobis transformation matrix}:\\
	Due to $C P_{k}^{-} C^{\top}+R>0,$ there exists an invertible matrix $F_{k} \in \mathbb{R}^{m \times m}$ such that
	$$F_{k} F_{k}^{\top}=\left(C P_{k}^{-} C^{\top}+R\right)^{-1}.$$
	Based on the raw measurement data $y_{k}$, the feedback data $C \hat{x}_{k}^{-}$ and $F_{k}$, the sensor calculates the innovation 
	\begin{equation}
	z_{k}=y_{k}-C \hat{x}_{k}^{-}
	\end{equation}
	where $z_{k} \sim \mathcal{N}(0, C P_{k}^{-} C^{\top}+R)$\cite{mo2013detecting}.
	Then the innovation's Mahalanobis transformation
	\begin{equation}
	\epsilon_{k} \triangleq F_{k}^{\top} z_{\mathrm{k}} \label{13}
	\end{equation}
	is sent to the event-based scheduler. Note that $\epsilon_{k} \sim \mathcal{N}(0, I_{m})$.
\end{itemize}

The event-based scheduler works as follows:
\begin{equation}
\gamma_{k}=\left\{\begin{array}{ll}
0, & \text { if }\left\|\epsilon_{k}\right\|_{\infty} \leq \beta \\
1, & \text { otherwise }
\end{array}\right. \label{14}
\end{equation}
where $\beta \geq 0$ is a fixed threshold.

\subsection{False-Data Detector}
At time $k$, the system utilizes a $\chi^{2}$ failure detector by checking the following hypothesis test:
\begin{equation}
g_{k}=z_{k}^{\top}\left(C P_{k}^{-} C^{\top}+R\right)^{-1} z_{k}=\epsilon_{k}^{\top}  \epsilon_{k}  \underset{H_{1}}{\overset{H_{0}}{\lessgtr}} \sigma \label{15}
\end{equation}
where the null hypothesis $H_{0}$ means that the system is operating normally, while the alternative hypothesis $H_{1}$ means that the system is under attack. When $H_{1}$ happens, an alarm will be launched. Note that $g_{k}$ is $\chi^{2}$ distributed with $m$ degrees of freedom.

Because $g_k \geq \left\|\epsilon_{k}\right\|_{\infty}^2$, it is necessary to require that $\beta < \sqrt{\sigma}$ to avoid the alarm being activated whenever the event-based scheduler is triggered in view of (\ref{14}).

\begin{rem}
	\label{rem:1}
	In practice, the threshold $\sigma$ could be designed corresponding to an acceptable false alarm rate $\Upsilon$, i.e., $\Pr(g_{k} \geq  \sigma) = \Upsilon$, see Definition 1 in the following.
\end{rem}

\subsection{Attack Model}
To launch attacks, the following assumptions are required over the communication channels.

\begin{assum}
	\label{ass:1}
	The attacker knows all the system parameters, e.g., $A$, $C$, $Q$, $R$ and $K$.
\end{assum}

\begin{assum}
	\label{ass:2}
	The attacker can modify the data transmitted through the forward and feedback channels.
\end{assum}

As shown in Fig. \ref{fig0}, we launch two attacks at both forward and feedback communication channels.

The converted innovation data $\epsilon_{k}$ is modified by the forward channel attacker as follows
\begin{equation}
\tilde{\epsilon}_{k}=\frac{1}{\mu_{k}} \epsilon_{k}+\delta_{k} \label{16}
\end{equation}
and
\begin{equation}
\tilde{z}_{k} = F_{k}^{-\top} \tilde{\epsilon}_{k} = \frac{1}{\mu_{k}} z_{k} + F_{k}^{-\top} \delta_{k}
\end{equation}
where $\tilde{\epsilon}_{k}$ and $\tilde{z}_{k}$ are the innovation modified by the attacker, $\mu_{k} > 1$ is a scaling parameter, and $\delta_{k} \in \mathbb{R}^{m}$ is a bias parameter independent of $\epsilon_{k}$. Our idea is to introduce a bias on the innovation to increase the communication rate of the event-based scheduler. Meanwhile, we manage the innovation with a linear compression to reduce the rate of being detected. It is noticed that the modified innovation has a non-zero mean and a smaller covariance, i.e.,  $\tilde{\epsilon}_{k} \sim \mathcal{N}(\delta_{k}, \frac{1}{\mu_{k}^2} I_{m})$.

The feedback channel attacker modifies the feedback data given as follows
\begin{equation}
C \hat{x}_{k}^{aa-} = C \hat{x}_{k}^{a-} + \alpha_{k} \label{17}
\end{equation}
where $\hat{x}_{k}^{a-}$ is the output of the remote estimator with the forward channel being attacked, $\hat{x}_{k}^{aa-}$ is the state estimation received by the sensor after the feedback attack, and $\alpha_{k}$ is the injected attack data to be designed. The purpose of this attack is to eliminate the influence of the forward channel attack so that $\epsilon_{k} \sim \mathcal{N}(0, I_{m})$ is kept.

Under the attack strategies (\ref{16}) and (\ref{17}), the data (\ref{15}) received by the false-data detector becomes
\begin{equation}
\tilde{g}_{k}=\tilde{\epsilon}_{k}^{\top} \tilde{\epsilon}_{k}. \label{18}
\end{equation}

Based on the above scheduler-pointed attack strategyies, we give the definition on successful attack.

\begin{defn}
	\label{def:1}
	The remote state estimation system with event-based sensor data scheduler in Fig. \ref{fig0} is called successfully attacked, if the following two conditions hold simultaneously:
	\begin{equation}
	\Pr(\|\tilde{\epsilon}_{k}\|_{\infty} \geq \beta) = \Pr(\gamma_{k}=1) \geq M \label{21}
	\end{equation}
	and
	\begin{equation}
	\Pr(\tilde{g}_{k} \geq  \sigma) \leq \Upsilon \label{22}
	\end{equation}
	where $0<M<1$ is a predesigned attack target, $\tilde{g}_{k}$ is the attacked $g_{k}$ in (\ref{15}), $\Upsilon$ is the false alarm rate of false-data detector. The set of all successful attacks is defined as $\mathscr{F}$.
\end{defn}

\begin{rem}
	\label{rem:2}
	In (\ref{16}), the coefficient of $\epsilon_{k}$ could be actually an arbitrary matrix for a generalized linear attack. But in our system, $\epsilon_{k} \sim \mathcal{N}(0, I_{m})$, so we can always find an equivalent diagonal matrix ${T}^{\rm diag}_{k}$ for an arbitrary matrix ${T}^{\rm arbi}_{k}$, i.e., ${T}^{\rm diag}_{k} \epsilon_{k}$ and ${T}^{\rm arbi}_{k} \epsilon_{k}$ are identically distributed. Thus, without loss of generality, the matrix $T_{k}$ is assumed to be diagonal. Moreover, a subset of $T_{k}$ which are uniform scaling parameters for $\epsilon_{k}$, as shown in (\ref{16}), is common and relatively concise for the operation of our compression on $\epsilon_{k}$.
\end{rem}

\subsection{Problem of Interest}
This paper is interested in constructing a successful attack, which degrades the effect of event-based scheduler to attack target $M$ while keeping the alarm rate of detector no more than the false alarm rate $\Upsilon$. Moreover, the evolution of the estimation performance after attack is also of our interests. These motivate our paper in the following sections.

\section{Scheduler-Pointed Attack Strategy}
This section studies the design of the two-channel scheduler-pointed FDI attack strategy. Firstly the existence of successful attacks is proven. The specific design method of successful attack strategies is then provided.

We rewrite the equation (\ref{18}) as
\begin{equation}
\tilde{g}_{k}=\tilde{\epsilon}_{k}^{\top} \tilde{\epsilon}_{k}=\frac{1}{\mu_{k}^{2}} \tilde{\epsilon}_{k}^{\top} \left( \frac{1}{\mu_{k}^2} I_{m} \right)^{-1} \tilde{\epsilon}_{k}.
\end{equation}
Note that $\tilde{g}_{k}$ is not $\chi^{2}$ distributed, which leads to the difficulties for our further analyse of (\ref{22}). Therefore, we can define a new variable
\begin{equation}
V_{k}=\tilde{\epsilon}_{k}^{\top} \left(\frac{1}{\mu_{k}^2} I_{m}\right)^{-1} \tilde{\epsilon}_{k}
\end{equation}
which is $\chi^{2}$ distributed. Then (\ref{22}) is equivalent to
\begin{equation}
\Pr(V_{k} \geq  \mu_{k}^2 \sigma) \leq \Upsilon. \label{19}
\end{equation}

Because of the Gaussianity of $\tilde{\epsilon}_{k}$ in (\ref{16}), the variable $V_{k}$ satisfies a noncentral $\chi^{2}$ distribution with $m$ degrees of freedom, i.e.,
\begin{equation}
V_{k} \sim \chi^{2} (m+\mu_{k}^{2} \phi_{k}^{2}, 2(m+2\mu_{k}^{2} \phi_{k}^{2}))
\end{equation}
where
\begin{equation}
\phi_{k} = \|\delta_{k}\|_{2}.
\end{equation}
We can further obtain the cumulative distribution function (CDF) of $V_{k}$ \cite{nuttal1975some}
\begin{equation}
F(x)=1-Q_{\frac{m}{2}}(\sqrt{\xi_{k}}, \sqrt{x}) \label{25}
\end{equation}
where
\begin{equation}
\xi_{k} = \mu_{k}^{2} \phi_{k}^{2}.
\end{equation}
Then the left part of (\ref{19}) can be written as
\begin{equation}
F(\mu_{k}^{2} \sigma) = \Pr(V_{k} \geq  \mu_{k}^2 \sigma) =  1-Q_{\frac{m}{2}}(\sqrt{\xi_{k}}, \sqrt{\mu_{k}^{2} \sigma}) \label{26}
\end{equation}
The $Q$ here represents the Marcum $Q$-function\cite{annamalai2009new}.

The following theorem states the existence of successful attacks.
\begin{thm}
	\label{thm:1}
	For any given $M,\Upsilon \in (0, 1)$, the feasible set $\mathscr{F} \neq \varnothing$.
\end{thm}
\begin{pf}
	We aims to find a pair of parameters $(\check{\mu}_{k}, \check{\delta}_{k}) \in \mathscr{F}$. Because $\beta < \sqrt{\sigma}$, there always exist vector $\check{\delta}_{k}$ satisfying $\left\|\check{\delta}_{k}\right\|_{\infty} > \beta$ and $\|\check{\delta}_{k}\|_{2} < \sqrt{\sigma}$. Let $\check{\psi}_{k} = \left\|\check{\delta}_{k}\right\|_{\infty}, \check{\phi}_{k} = \|\check{\delta}_{k}\|_{2}.$
	
	For (\ref{21}), one obtains that
	$$
	\begin{aligned}
	\Pr(\|\tilde{\epsilon}_{k}\|_{\infty} \geq \beta) &\geq \Pr(|\tilde{\epsilon}_{ki}| \geq \beta)\\
	&= \Pr(|\frac{\epsilon_{ki}}{\mu_{k}}+\check{\delta}_{ki}| \geq \beta)
	\end{aligned}
	$$
	for any $i = 1,\cdots,m$, where $\tilde{\epsilon}_{ki}$, $\epsilon_{ki}$ and $\check{\delta}_{ki}$ are the $i$-th entry of $\tilde{\epsilon}_{i}$, $\epsilon_{k}$ and $\check{\delta}_{k}$, respectively. Assuming $|\check{\delta}_{kj}| = \check{\psi}_{k}$, we obtain
	$$
	\begin{aligned}
	\Pr(\|\tilde{\epsilon}_{k}\|_{\infty} \geq \beta) &\geq \Pr(|\frac{\epsilon_{kj}}{\mu_{k}}+\check{\psi}_{k}| \geq \beta)\\
	&\geq \Pr(\frac{\epsilon_{kj}}{\mu_{k}}+\check{\psi}_{k} \geq \beta)\\
	&= \Pr[\epsilon_{kj} \geq \mu_{k}(\beta-\check{\psi}_{k})].
	\end{aligned}
	$$
	Because $\epsilon_{kj}$ satisfies the standard Gaussian distribution and $\beta < \check{\psi}_{k}$, there is always a solution $\mu_{k} = \tau_{1}$ to
	\begin{equation}
	\Pr[\epsilon_{kj} \geq \mu_{k}(\beta-\check{\psi}_{k})] \geq M
	\end{equation}
	and it is obvious that all $\mu_{k} \geq \tau_{1}$ meet (\ref{21}).
	
	Then considering (\ref{22}), we have
	$$
	\begin{aligned}
	\Pr(\tilde{g}_{k} \geq \sigma) &= \Pr(\tilde{\epsilon}_{k}^{\top} \tilde{\epsilon}_{k} \geq \sigma)\\
	&= \Pr \bigg [\sum_{i=1}^{m}(\frac{\epsilon_{ki}}{\mu_{k}}+\check{\delta}_{ki})^2 \geq \sigma \bigg]\\
	&= \Pr \bigg [\sum_{i=1}^{m}(\frac{\epsilon_{ki}^2}{\mu_{k}^2} + \frac{2\epsilon_{ki}\check{\delta}_{ki}}{\mu_{k}}) \geq \sigma-\check{\phi}_{k}^2 \bigg ] \\
	&= 1 - \Pr \bigg [\sum_{i=1}^{m}(\frac{\epsilon_{ki}^2}{\mu_{k}^2} + \frac{2\epsilon_{ki}\check{\delta}_{ki}}{\mu_{k}}) \leq \sigma-\check{\phi}_{k}^2 \bigg ]
	\end{aligned}
	$$
	where
	$$
	\begin{aligned}
	\Pr \bigg [\sum_{i=1}^{m}(\frac{\epsilon_{ki}^2}{\mu_{k}^2} + \frac{2\epsilon_{ki}\check{\delta}_{ki}}{\mu_{k}}) \leq \sigma-\check{\phi}_{k}^2 \bigg]& \\
	\geq \Pr \bigg [\Big(\sum_{i=1}^{m} \epsilon_{ki}^2 \leq \frac{\mu_{k}^2(\sigma-\check{\phi}_{k}^2)}{2} \Big)& \\
	\cap \Big(\sum_{i=1}^{m} 2\epsilon_{ki}\check{\delta}_{ki} &\leq \frac{\mu_{k}(\sigma-\check{\phi}_{k}^2)}{2} \Big) \bigg].
	\end{aligned}
	$$
	Similarly, because $\sum_{i=1}^{m} \epsilon_{ki}^2$ is $\chi^2$ distributed, $\sum_{i=1}^{m} 2\epsilon_{ki}\check{\delta}_{ki}$ is Gaussian distributed and $\sigma>\check{\phi}_{k}^2$, there is always a solution $\mu_{k} = \tau^{\prime}$ to
	\begin{equation}
	\Pr \bigg[\sum_{i=1}^{m} \epsilon_{ki}^2 \leq \frac{\mu_{k}^2(\sigma-\check{\phi}_{k}^2)}{2} \bigg] \geq 1-\frac{\Upsilon}{2} \label{27}
	\end{equation}
	and a solution $\mu_{k} = \tau^{\prime\prime}$ for
	\begin{equation}
	\Pr \bigg[\sum_{i=1}^{m} 2\epsilon_{ki}\check{\delta}_{ki} \leq \frac{\mu_{k}(\sigma-\check{\phi}_{k}^2)}{2} \bigg] \geq 1-\frac{\Upsilon}{2} \label{28}
	\end{equation}
	and for all $\mu_{k} \geq \tau_{2} = \max \{\tau^{\prime}, \tau^{\prime\prime}\}$, the inequalities in (\ref{27}) and (\ref{28}) are true simultaneously. Hence, we obtain
	$$
	\begin{aligned}
	\Pr \bigg [\Big(\sum_{i=1}^{m} \epsilon_{ki}^2 \leq \frac{\mu_{k}^2(\sigma-\check{\phi}_{k}^2)}{2} \Big)& \\
	\cap \Big(\sum_{i=1}^{m} 2\epsilon_{ki}\check{\delta}_{ki} &\leq \frac{\mu_{k}(\sigma-\check{\phi}_{k}^2)}{2} \Big) \bigg] \geq 1-\Upsilon
	\end{aligned}
	$$
	namely
	$$
	Pr(\tilde{g}_{k} \geq \sigma) \leq \Upsilon.
	$$
	
	Now let $\check{\mu}_{k} \geq \max \{\tau_{1}, \tau_{2}\}$, we obtain a feasible attack parameter pair $(\check{\mu}_{k}, \check{\delta}_{k}) \in \mathscr{F}$. $\hfill\blacksquare$
\end{pf}

 \begin{rem}
	\label{rem:3}
	Note that Theorem \ref{thm:1} implies no matter how large the predesigned target percentage $M$ is, the attackers can always launch a successful attack. That is, the event-based scheduler can be deemed to be completely invalidated when an $M$ is chosen to be large enough.
\end{rem}

Next, to keep the Gaussian distribution of innovation $z_{k}$, the feedback channel attack (\ref{17}) is designed.

Under the attack (\ref{16}), the Kalman filter in (\ref{3})-(\ref{5}) becomes
\begin{equation}
\left\{\begin{aligned}
\hat{x}_{k}^{a-}&=A \hat{x}_{k-1}^{a}  \\
\hat{x}_{k}^{a}&=\hat{x}_{k}^{a-}+\gamma_{k} K_{k} F_{k}^{-\top} \tilde{\epsilon}_{k}
\end{aligned}\right. \label{30}
\end{equation}
where $\hat{x}_{k}^{a-}$ and $\hat{x}_{k}^{a}$ are the \textit{priori} and \textit{posteriori} MMSE estimates after attack (\ref{16}), respectively. The attack effect is defined as
\begin{align}
\tilde{x}_{k}^{-} &\triangleq \hat{x}_{k}^{a-} - \hat{x}_{k}^{-}\\
\tilde{x}_{k} &\triangleq \hat{x}_{k}^{a} - \hat{x}_{k}. \label{32}
\end{align}
Then one has that
\begin{equation}
\begin{aligned}
\tilde{x}_{k}^{-} &= A(\hat{x}_{k-1}^{a} - \hat{x}_{k-1}) = A \tilde{x}_{k-1}
\end{aligned}
\end{equation}
and
\begin{equation}
\begin{aligned}
\tilde{x}_{k} &= \hat{x}_{k}^{a-} - \hat{x}_{k}^{-} + \gamma_{k} K_{k} F_{k}^{-\top} \tilde{\epsilon}_{k} - \gamma_{k} K_{k} z_{k} \\
&= \tilde{x}_{k}^{-} + \gamma_{k} K_{k} F_{k}^{-\top} (\frac{1}{\mu_{k}} \epsilon_{k}+\delta_{k}) - \gamma_{k} K_{k} z_{k} \\
&= \tilde{x}_{k}^{-} + (\frac{\gamma_{k}}{\mu_{k}} - \gamma_{k}) K_{k} z_{k} + \gamma_{k} K_{k} F_{k}^{-\top} \delta_{k}.
\end{aligned} \label{33}
\end{equation}
According to the steady-state assumption $\hat{x}_{0}^{a} = \hat{x}_{0}$, we have $\tilde{x}_{0} = 0$,
then we can calculate $\tilde{x}_{k}$ and $\tilde{x}_{k}^{-}$ at any time $k$ by iteration.

Next, the innovation after our two attacks (\ref{16}) and (\ref{17}), denoted by $z_{k}^{a}$, is expressed as
\begin{equation}
\begin{aligned}
z_{k}^{a} &\triangleq y_{k} - C \hat{x}_{k}^{aa-}\\
&=y_{k}-(C \hat{x}_{k}^{a-} + \alpha_{k}) \\
&=y_{k}-C (\hat{x}_{k}^{-} + \tilde{x}_{k}^{-}) - \alpha_{k}\\
&=y_{k}-C \hat{x}_{k}^{-} - C \tilde{x}_{k}^{-} - \alpha_{k}\\
&=z_{k} - C \tilde{x}_{k}^{-} - \alpha_{k}.
\end{aligned}
\end{equation}
In order to ensure that $z_{k}^{a}$ follows the same distribution as $z_{k}$, the feedback channel attack can be designed as
\begin{equation}
\alpha_{k} = - C \tilde{x}_{k}^{-}\label{34}
\end{equation}

Based on the two-channel scheduler-pointed attack discussed above, it can be observed that the successful attack strategies can be persistently launched against the event-based remote estimation system. Next, we will consider how to achieve such attacks.

According to Theorem \ref{thm:1}, the attackers can always launch a successful attack for a pair of given performance index $(M, \Upsilon)$ as long as a large enough scaling parameter $\mu_{k}$ is chosen. Naturally, the infimum of $\mu_{k}$, by which the feasible set of $\mu_{k}$ can be obtained, is worth investigating. For any fixed $\mu_{k}$, we can simplify constraints (\ref{21}) and (\ref{22}) to the single variable form that depends on $\delta_{k}$ only. Then a successful attack parameter pair $(\check{\mu}_{k}, \check{\delta}_{k})$ can be designed. We first give an assumption.

\begin{assum}
	\label{ass:3}
	The bias parameter $\delta_{k}$ has the form of
	$$
	\delta_{k}=\left[\begin{array}{llll}
	\bar{\delta}_{k} & 0 & \cdots & 0
	\end{array}\right].
	$$
\end{assum}

\begin{rem}
	\label{rem:4}
	Assumption \ref{ass:3} is reasonable because the event-based scheduler is triggered according to the infinity norm as shown in (\ref{15}). This assumption does not introduce any conservativeness and is more economical. In addition, the $\bar{\delta}_{k}$ can be put at any position without loss of generality due to the independence of the $m$ elements of $\epsilon_{k}$.
\end{rem}

For satisfying Definition 1, according to (\ref{19}), (\ref{26}) and the Gaussianity of $\tilde{\epsilon}_{k}$, the optimization problem about $\mu_{k}$ can be formulated as follows:
\begin{prob}
	\label{pro:1}
	\begin{equation}
	\begin{gathered}
	\min _{\mu_{k} \geq 1} \mu_{k} \\
	\text {\rm s.t. }
	\left\{\begin{aligned}
	Q_{\frac{m}{2}}(\mu_{k} \bar{\delta}_{k}, \mu_{k} \sqrt{\sigma}) &\leq \Upsilon \\
	\mu_{k} (\bar{\delta}_{k} - \beta) &\geq \Psi
	\end{aligned}\right.
	\end{gathered} \label{35}
	\end{equation}
	where $\Psi$ is defined as the confidence interval of standard Gaussian distribution determined by
	the percentage $M$, i.e.,
	\begin{equation}
	\frac{1}{2}\left[1+\operatorname{erf}\left(\frac{\Psi}{\sqrt{2}}\right)\right] = M. \label{36}
	\end{equation}
\end{prob}

This is a nonlinear programming problem which is difficult to solve by traditional methods. Luckily, the solution to this problem can be obtained from the following theorem and one can complete its proof using the monotonicity of functions.

\begin{thm}
	\label{thm:2}
	The attack parameter pair $(\mu_{k}^{\ast}, \bar{\delta}_{k}^{\ast})$ is an optimal solution to (\ref{35}), only if they satisfy the following two constraints
	\begin{equation}
	\left\{\begin{aligned}
	Q_{\frac{m}{2}}(\mu_{k}^{\ast} \bar{\delta}_{k}^{\ast}, \mu_{k}^{\ast} \sqrt{\sigma}) &= \Upsilon   \\
	\mu_{k}^{\ast} (\bar{\delta}_{k}^{\ast} - \beta) &= \Psi. \\
	\end{aligned}\right. \label{37}
	\end{equation}
\end{thm}

To prove theorem \ref{thm:2}, we need two preliminary results.

\begin{lem}~\cite{sun2008inequalities}
	\label{lem:1}
	For independent random variables $X$ and $Y$ with nonnegative support, we have
	\begin{equation}
	\Pr(X+Y \geq c)>\Pr(X \geq c), \quad \text { for any } c>0
	\end{equation}
\end{lem}

\begin{lem}
	\label{lem:2}
	The Marcum $Q$-function $Q_{\nu}(a, b)$ is strictly increasing in $a$ for all $a \geq 0$ and $b>0, \nu>0$, and strictly decreasing in $b$ for all $a\geq 0, b\geq 0$ and $ \nu>0$.
\end{lem}
\begin{pf}
	Let $X \sim \chi_{\nu, a_{1}}^{2}$ and $Y \sim \chi_{0, a_{2}}^{2}$, then $X+Y \sim \chi_{\nu, a_{1}+a_{2}}^{2}$. Then it follows from Lemma \ref{lem:1} that
	\begin{equation}
	Q_{\frac{\nu}{2}}\left(\sqrt{a_{1}+a_{2}}, \sqrt{b}\right)>Q_{\frac{\nu}{2}}\left(\sqrt{a}_{1}, \sqrt{b}\right)
	\end{equation}
	for all $a_{1} \geq 0$ and $a_{2}, b, \nu>0$.
	
	Moreover, it can be verified that $F(x)$ in (\ref{25}) is strictly increasing with respect to $x$ on $[0, \infty)$ because of the monotonicity of CDF. Hence, the strictly decreasing in $b$ holds. The proof is now completed. $\hfill\blacksquare$
\end{pf}

Now we are ready to prove Theorem \ref{thm:2}.
\begin{pf}
	(\textbf{Theorem \ref{thm:2}}) Let us prove it by contradiction. Assume that $(\mu_{k}^{\ast}, \bar{\delta}_{k}^{\ast})$ is an optimal solution to Problem \ref{pro:1} while it does not satisfy (\ref{37}). Such a condition can be divided into the following two cases:
	\begin{itemize}
		\item[1)] case one:
		\begin{equation}
		Q_{\frac{m}{2}}(\mu_{k}^{\ast} \bar{\delta}_{k}^{\ast}, \mu_{k}^{\ast} \sqrt{\sigma}) < \Upsilon.
		\end{equation}
		From Lemma \ref{lem:2}, there always exists a $\Delta$ satisfying $(\mu_{k}^{\ast}-1) \sqrt{\sigma}>\Delta>0$, such that
		$$Q_{\frac{m}{2}}(\mu_{k}^{\ast} \bar{\delta}_{k}^{\ast}, \mu_{k}^{\ast} \sqrt{\sigma}-\Delta) < \Upsilon.$$
		We denote
		$$
		\begin{aligned}
		\mu_{k}^{\prime} &= \mu_{k}^{\ast} - \frac{\Delta}{\sqrt{\sigma}} \\
		\bar{\delta}_{k}^{\prime} &= \frac{\mu_{k}^{\ast} \bar{\delta}_{k}^{\ast}}{\mu_{k}^{\prime}}.
		\end{aligned}
		$$
		It can be verified that attack parameter pair $(\mu_{k}^{\prime}, \bar{\delta}_{k}^{\prime})$ satisfies constraints (\ref{35}). Because $\mu_{k}^{\prime}<\mu_{k}^{\ast}$, $\mu_{k}^{\ast}$ is not optimal, which contradicts the assumption.
		\item[2)] case two:
		\begin{equation}
		Q_{\frac{m}{2}}(\mu_{k}^{\ast} \bar{\delta}_{k}^{\ast}, \mu_{k}^{\ast} \sqrt{\sigma}) = \Upsilon
		\end{equation}
		and
		\begin{equation}
		\mu_{k}^{\ast} (\bar{\delta}_{k}^{\ast} - \beta) > \Psi \label{42}.
		\end{equation}
		Consider the inequality (\ref{42}), there always exists a small enough $\Delta>0$, such that
		$$\mu_{k}^{\ast} (\bar{\delta}_{k}^{\ast} - \beta) - \Delta > \Psi$$
		then we denote
		$$
		\begin{aligned}
		\mu_{k}^{\prime} &= \mu_{k}^{\ast}\\
		\bar{\delta}_{k}^{\prime} &= \bar{\delta}_{k}^{\ast} - \frac{\Delta}{\mu_{k}^{\ast}}.
		\end{aligned}
		$$
		The attack parameter pair $(\mu_{k}^{\prime}, \bar{\delta}_{k}^{\prime})$ satisfies (\ref{35}). Moreover, although $\mu_{k}^{\prime}$ remains unchanged, we have $Q_{\frac{m}{2}}(\mu_{k}^{\prime} \bar{\delta}_{k}^{\prime}, \mu_{k}^{\prime} \sqrt{\sigma}) < Q_{\frac{m}{2}}(\mu_{k}^{\ast} \bar{\delta}_{k}^{\ast}, \mu_{k}^{\ast} \sqrt{\ast}) = \Upsilon$ due to the decrease of $\bar{\delta}_{k}^{\prime}$. Then we notice that $(\mu_{k}^{\prime}, \bar{\delta}_{k}^{\prime})$ is also a solution satisfying case one. This means a better solution can be found as above, which contradicts the assumption.	The proof is now completed. $\hfill\blacksquare$
	\end{itemize}
\end{pf}

Finally, we present an efficient method for solving the equations in (\ref{37}). Marcum $Q$-function with the unknown $\mu_{k}$ is difficult to solve because it is of an integral form as shown in its definition \cite{marcum1950table}. In \cite{sun2010monotonicity}, the Marcum $Q$-function $Q_{\nu}(a, b)$ was written as a closed-form expression
\begin{equation}
\begin{aligned}
Q_{\nu}(a, b)=& \frac{1}{2} \operatorname{erfc}\left(\frac{b+a}{\sqrt{2}}\right)+\frac{1}{2} \operatorname{erfc}\left(\frac{b-a}{\sqrt{2}}\right) \\
&+\frac{1}{a \sqrt{2 \pi}} \sum_{k=0}^{\nu-1.5} \frac{b^{2 k}}{2^{k}} \sum_{q=0}^{k} \frac{(-1)^{q}(2 q) !}{(k-q) ! q !} \\
& \times \sum_{i=0}^{2 q} \frac{1}{(a b)^{2 q-i} i !}\left[(-1)^{i} e^{-\frac{(b-a)^{2}}{2}}-e^{-\frac{(b+a)^{2}}{2}}\right] \\
&\qquad\qquad\qquad\qquad\qquad\qquad a>0,  b \geq 0 \label{46}
\end{aligned}
\end{equation}
where $\nu$ is an odd multiple of 0.5. For the case when $\nu$ is an even multiple of 0.5, the average of $Q_{\nu-0.5}(a, b)$ and $Q_{\nu+0.5}(a, b)$ was shown to be a good approximation \cite{li2006computing}.

This representation of Marcum $Q$-function involves only the exponential and $\operatorname{erfc}$ functions. Well-designed mathematical library functions for the computation of $\operatorname{erfc}$ functions are available in most algorithmic languages (such as Fortran or MATLAB). Hence, one can easily solve the equations (\ref{37}), both numerically and analytically. Therefore, a relatively integrated solving method for Problem \ref{pro:1} is given, and the performance analysis of this scheduler-pointed FDI attack strategy will be shown in the following section.

\begin{rem}
	\label{rem:5}
	Notice that design of the attack strategy parameters $\mu_{k}$ and $\bar{\delta}_{k}$ is only associated with the distribution of $\epsilon_{k}$, which is identically $\mathcal{N}(0, I_{m})$. This means that the optimal solution to Problem \ref{pro:1} is independent of time index $k$. Thus, we denote $\delta_{k}$, $\mu_{k}$ and $\bar{\delta}_{k}$ as $\delta$, $\mu$ and $\bar{\delta}$ for simplicity after this.
\end{rem}

\section{Performance Analysis}
In this section, we derive the evolution of the mean and covariance of the estimation error at the remote estimator during an attack. The system performance degradation is also quantified and analyzed.

According to Theorem \ref{thm:1} and Remark \ref{rem:3}, when the attack target $M$ is close to $1$, the trigger rate of the event-based scheduler is almost $1$, i.e.,
\begin{equation}
\lim_{M \to 1} [\Pr(\gamma_{k}=1)] = 1 \label{47}
\end{equation}
In practice, the attackers always tend to launch a more powerful attack. Therefore, the following assumption is reasonable.
\begin{assum}
	\label{ass:4}
	$\Pr(\gamma_{k}=1) = 1$.
\end{assum}
Unless specifically mentioned, our analysis in this section is based on this assumption. This assumption leads to a very simple form of performance analysis by degrading the remote estimator (\ref{3})-(\ref{5}) to a standard Kalman filter. Without loss of generality, it is assumed that the estimator has converged to steady-state after time $k_{a}$\cite{mo2013detecting} and let
$$
\begin{aligned}
P &\triangleq \lim _{k \rightarrow \infty} P^{-}_{k}\\
K &\triangleq P C^{\top}\left(C P C^{\top}+R\right)^{-1}
\end{aligned}
$$

From iteration (\ref{32}) and (\ref{33}), one has
$$
\hat{x}_{k}^{a}-\hat{x}_{k} = A(\hat{x}_{k-1}^{a}-\hat{x}_{k-1}) + (\frac{1}{\mu} - 1) K z_{k} + K F^{-\top} \delta
$$
then
\begin{equation}
\mathbf{E}(\hat{x}_{k}^{a}-\hat{x}_{k}) = A\mathbf{E}(\hat{x}_{k-1}^{a}-\hat{x}_{k-1}) + K F^{-\top} \delta.
\end{equation}
If $A$ is unstable, one has $\lim _{k \rightarrow \infty} \mathbf{E}(\hat{x}_{k}^{a}-\hat{x}_{k})=\infty$. If $A$ is stable, we can verify that $\hat{x}_{k}^{a}-\hat{x}_{k}$ converges to a Gaussian distribution with mean $(I_{n}-A)^{-1} K F^{-\top} \delta$. Note that $(I_{n}-A)^{-1}$ exists due to the stability of $A$. Without loss of generality, we assume that for  $k \geq k_{a}$
\begin{equation}
\mathbf{E}(\hat{x}_{k}^{a}-\hat{x}_{k}) = (I_{n}-A)^{-1} K F^{-\top} \delta. \label{50}
\end{equation}
Substituting (\ref{50}) into (\ref{33}), we have
\begin{equation}
\mathbf{E}(\hat{x}_{k}^{a-}-\hat{x}_{k}^{-}) = A (I_{n}-A)^{-1} K F^{-\top} \delta. \label{50-2}
\end{equation}

The following theorem summarizes the evolution of the estimation error covariance at the remote estimator under attack.
\begin{thm}
	\label{thm:3}
	Under Assumption \ref{ass:4}, the estimation error covariance of the remote estimator follows the recursion:
	\begin{equation}
	\begin{aligned}
	P_{k+1}^{a} =& A P_{k}^{a} A^{\top} + Q -(\frac{2}{\mu}-\frac{1}{\mu^{2}})PC^{\top}S^{-1}CP
	\end{aligned} \label{51}
	\end{equation}
	where $S = C P C^{\top} + R$, $k \geq k_{a}$.
\end{thm}
\begin{pf}
	Substituting the process model (\ref{1}), (\ref{2}) and the innovation (\ref{13}), (\ref{16}) into the estimator (\ref{30}), we have
	$$
	\begin{aligned}
	\hat{x}_{k}^{a-}-x_{k} &= A (\hat{x}_{k-1}^{a}-x_{k-1}) - w_{k-1} \\
	\hat{x}_{k}^{a}-x_{k} &= \hat{x}_{k}^{a-}-x_{k} + K F^{-\top} \tilde{\epsilon}_{k}\\
	&= \hat{x}_{k}^{a-}-x_{k} + \frac{1}{\mu}Kz_{k} + K F^{-\top} \delta.
	\end{aligned}
	$$
	We denote
	$$\mathscr{E} \triangleq (I_{n}-A)^{-1} K F^{-\top} \delta. $$
	Notice that $\mathbf{E}(\hat{x}_{k}) = \mathbf{E}(\hat{x}_{k}^{-}) = \mathbf{E}(x_{k})$, by (\ref{50}) and (\ref{50-2}), we obtain
	$$\mathbf{E}(\hat{x}_{k}^{a}-x_{k}) = \mathscr{E}$$
	$$\mathbf{E}(\hat{x}_{k}^{a-}-x_{k}) = A \mathscr{E}.$$
	We have an equation that
	\begin{equation}
	\mathscr{E} - A \mathscr{E} = (I_{n}-A) \mathscr{E} = K F^{-\top} \delta.
	\end{equation}
	
	Then the error covariance at the remote estimator can be expressed as
	\begin{equation}
	\begin{aligned}
	P_{k }^{a-} =& \mathbf{E}[\hat{x}_{k}^{a-}-x_{k} - A\mathscr{E})(\hat{x}_{k}^{a-}-x_{k} - A\mathscr{E})^{\top}] \\
	=& \mathbf{E}[(A(\hat{x}_{k-1}^{a}-x_{k}-\mathscr{E}) - w_{k-1})(*)^{\top}] \\
	=& A P_{k-1}^{a} A^{\top} + Q \\
	P_{k }^{a} =& \mathbf{E}[\hat{x}_{k}^{a}-x_{k} - \mathscr{E})(\hat{x}_{k}^{a}-x_{k} - \mathscr{E})^{\top}] \\
	=& \mathbf{E}[(\hat{x}_{k}^{a-}-x_{k} - A\mathscr{E} + \frac{1}{\mu}Kz_{k})(*)^{\top}] \\
	=& P_{k }^{a-} + \frac{1}{\mu^2}KSK^{\top} + \frac{1}{\mu}\mathbf{E}[(\hat{x}_{k}^{a-}-x_{k} -  A\mathscr{E})z_{k}^{\top}K^{\top}] \\
	&+ \frac{1}{\mu}\mathbf{E}[Kz_{k}(\hat{x}_{k}^{a-}-x_{k} - A\mathscr{E})^{\top}].
	\end{aligned} \label{53}
	\end{equation}
	
	To calculate the last two terms of (\ref{53}), first we obtain
	\begin{equation}
	\begin{aligned}
	\hat{x}_{k}^{a-}&-x_{k} - A\mathscr{E} \\
	=& A(\hat{x}_{k-1}^{a}-x_{k-1}) - A\mathscr{E} - w_{k-1} \\
	=& A(\hat{x}_{k-1}^{a-} + \frac{1}{\mu}Kz_{k-1} + K F^{-\top} \delta) - Ax_{k-1} - A\mathscr{E} - w_{k-1} \\
	=& A(\hat{x}_{k-1}^{a-} - x_{k-1} - A\mathscr{E}) - \frac{1}{\mu}AKz_{k-1} - w_{k-1} \\
	=& A^{k-k_{a}}(\hat{x}_{k_{a}}^{-} - x_{k_{a}} - A\mathscr{E}) - \sum_{i=0}^{k-1-k_{a}} A^{i} w_{k-1-i} \\
	&+ \frac{1}{\mu} \sum_{i=0}^{k-1-k_{a}} A^{i+1} K z_{k-1-i}. \label{54}
	\end{aligned}
	\end{equation}
	where the last equality follows from the steady-state assumption $\hat{x}_{k_{a}}^{a-} = \hat{x}_{k_{a}}^{-}$.
	
	On the other hand, the innovation $z_{k}$ is given by seeing Lemma 1 in \cite{guo2018worst},
	\begin{equation}
	\begin{aligned}
	z_{k} =& C [A(I_{n}-K C)]^{k-k_{a}}(x_{k_{a}}-\hat{x}_{k_{a}}^{-}) \\
	&+ \sum_{i=0}^{k-1-k_{a}} C[A(I_{n}-K C)]^{i} w_{k-1-i} + \mathscr{V}
	\end{aligned}
	\end{equation}
	where $\mathscr{V} = v_{k} - \sum_{i=0}^{k-1-k_{a}} C[A(I_{n}-K C)]^{i} AKv_{k-1-i}$, which is independent of the first two terms in (\ref{54}). Notice that innovation $z_{k}$ is Gaussian i.i.d.. Now the undetermined terms of (\ref{53}) can be calculated as
	\begin{equation}
	\begin{aligned}
	\mathbf{E}[&(\hat{x}_{k}^{a-}-x_{k} - A\mathscr{E}  )z_{k}^{\top}K^{\top}] \\
	=& \mathbf{E}\Bigg[  \Bigg \{  A^{k-k_{a}} (\hat{x}_{k_{a}}^{-} - x_{k_{a}} - A\mathscr{E}) - \sum_{i=0}^{k-1-k_{a}} A^{i} w_{k-1-i}\Bigg \}  \\
	&\qquad \cdot \Bigg \{ C [A(I_{n}-K C)]^{k-k_{a}}(x_{k_{a}}-\hat{x}_{k_{a}}^{-}) \\
	&\quad \qquad + \sum_{i=0}^{k-1-k_{a}} C[A(I_{n}-K C)]^{i} w_{k-1-i} \Bigg \}^{\top}  K^{\top}  \Bigg] \\
	=& \Bigg \{A^{k-k_{a}} \mathbf{E}\Big[(\hat{x}_{k_{a}}^{-} - x_{k_{a}} - A\mathscr{E})(x_{k_{a}}-\hat{x}_{k_{a}}^{-})^{\top}\Big] \\
	&\qquad\qquad\qquad\qquad\qquad \cdot [(I_{n}-KC)^{\top} A^{\top}]^{k-k_{a}} \\
	& - \sum_{i=0}^{k-1-k_{a}} A^{i}\mathbf{E}[w_{k-1-i} w_{k-1-i}^{\top}][(I_{n}-KC)^{\top} A^{\top}]^{i} \Bigg \} C^{\top} K^{\top} \\
	=& -\Bigg \{A^{k-k_{a}} P [(I_{n}-KC)^{\top} A^{\top}]^{k-k_{a}} \\
	& \ \quad + \sum_{i=0}^{k-1-k_{a}} A^{i}Q[(I_{n}-KC)^{\top} A^{\top}]^{i} \Bigg \}C^{\top} K^{\top}.
	\end{aligned} \label{57}
	\end{equation}
	Note that $P$ is the unique positive semi-definite fixed point of $h \circ \tilde{q}$ \cite{guo2018worst}, i.e., $ P=(h \circ \tilde{q})^{k-k_{a}}(P)$, (\ref{57}) can be further simplified as
	\begin{equation}
	\mathbf{E}[(\hat{x}_{k}^{a-}-x_{k} - A\mathscr{E}  )z_{k}^{\top}K^{\top}] = - PC^{\top} K^{\top} \label{59}
	\end{equation}
	and similarly
	\begin{equation}
	\mathbf{E}[Kz_{k}(\hat{x}_{k}^{a-}-x_{k} - A\mathscr{E})^{\top}] = -KCP. \label{60}
	\end{equation}
	
	Substituting (\ref{59}) and (\ref{60}) into (\ref{53}), the estimation error covariance becomes
	\begin{equation}
	\begin{aligned}
	P_{k }^{a} =& P_{k }^{a-} + \frac{1}{\mu^2}KSK^{\top} - \frac{1}{\mu}PC^{\top} K^{\top} - \frac{1}{\mu}KCP \\
	=& A P_{k-1}^{a} A^{\top} + Q + \frac{1}{\mu^2}PC^{\top}S^{-1}CP \\
	&- \frac{1}{\mu}PC^{\top}S^{-1}CP - \frac{1}{\mu}PC^{\top}S^{-1}CP \\
	=& A P_{k-1}^{a} A^{\top} + Q - (\frac{2}{\mu}-\frac{1}{\mu^{2}})PC^{\top}S^{-1}CP.
	\end{aligned} \label{61}
	\end{equation}
	The proof is now completed. $\hfill\blacksquare$
\end{pf}

According to Theorem \ref{thm:3}, the remote estimator turns to be biased and possesses a larger error covariance after our attack strategy, i.e., $\hat{x}_{k}^{a}-x_{k}\sim \mathcal{N}(\mathscr{E}, P_{k }^{a})$. The following properties of $P_{k }^{a}$ can be obtained:
\begin{cor}
	\label{cor:1}
	$P_{k }^{a}$ is monotonically increasing for $\mu \in [1, +\infty)$.
\end{cor}

\begin{cor}
	\label{cor:2}
	There is always $P_{k }^{a} \geq P_{k }$ and the equality holds if and only if the linear attack is off, i.e., $\mu = 1$.
\end{cor}

\begin{cor}
	\label{cor:3}
	$P_{k }^{a}$ converges to the open loop form of the remote estimator
	\begin{equation}
	P_{k }^{o} = h(P_{k-1}^{o}) = A P_{k-1}^{o} A^{\top} + Q \label{65}
	\end{equation}
	when $\mu$ approaches infinity.
\end{cor}

\section{Numerical Simulation}
In this section, numerical simulations are provided to illustrate the effectiveness of our scheduler-pointed FDI attack.

Consider a stable system with randomly generated parameters
$$A=
\left[
\begin{array}{ccc}
0.5944 & -0.1203 & -0.4302\\
0.0017 & 0.7902 & -0.0747\\
0.0213 & 0.8187 & 0.1436
\end{array}
\right] $$
$$C =
\left[
\begin{array}{ccc}
0.1365 & 0.8939 & 0.2987\\
0.0118 & 0.1991 & 0.6614
\end{array}
\right]   $$
with covariance $Q = 0.01I_{3}$ and $R = 0.1I_{3}$. Assuming the false alarm rate of the system build-in $\chi^{2}$ detector is no more than $1\%$, i.e., $\Upsilon = 0.01$. The fixed threshold of the event-based scheduler is set to be $\beta = 1.4$ and the attack target $M=99.87\%$. By (\ref{36}) one can calculate that $\Psi = 3$.  Then it can be verified that $\sigma = 11.34$ from the $\chi^2$ table with $3$ degrees of freedom.

Using the parameters above, the optimal solution of Problem (\ref{35}) is found by solving (\ref{37}) and (\ref{46}) to be
\begin{equation}
(\mu^{\ast}, \bar{\delta}^{\ast}) = (2.7705, 2.4828). \label{62}
\end{equation}
In fact, with this optimal solution, we have obtained the whole feasible solution set of our attack strategy because $\mu^{\ast}$ is the infimum of all feasible $\mu$. For any given $\mu \geq \mu^{\ast}$, the feasible set of $\delta$ can be obtained easily by solving Problem \ref{pro:1}. Our following simulations are based on (\ref{62}) because it is obvious that a bigger $\mu$ will lead to more efficient attack effects.

\begin{figure}[!t]
	\centering
	\includegraphics[width=3.5in]{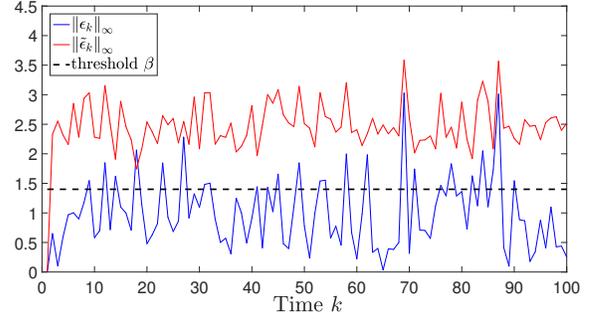}
	\caption{The triggering of event-based scheduler.}
	\label{fig1}
\end{figure}

The simulation result of the event-based scheduler is shown in Fig. \ref{fig1}. Specifically, the communication rate without attack is $29.69\%$. Under our two-channel scheduler-pointed attack, the rate is upgraded to $99.98\%$, which satisfies our attack target. The event-based estimator is now almost equivalent to the standard Kalman filter.

\begin{figure}[!t]
	\centering
	\includegraphics[width=3.5in]{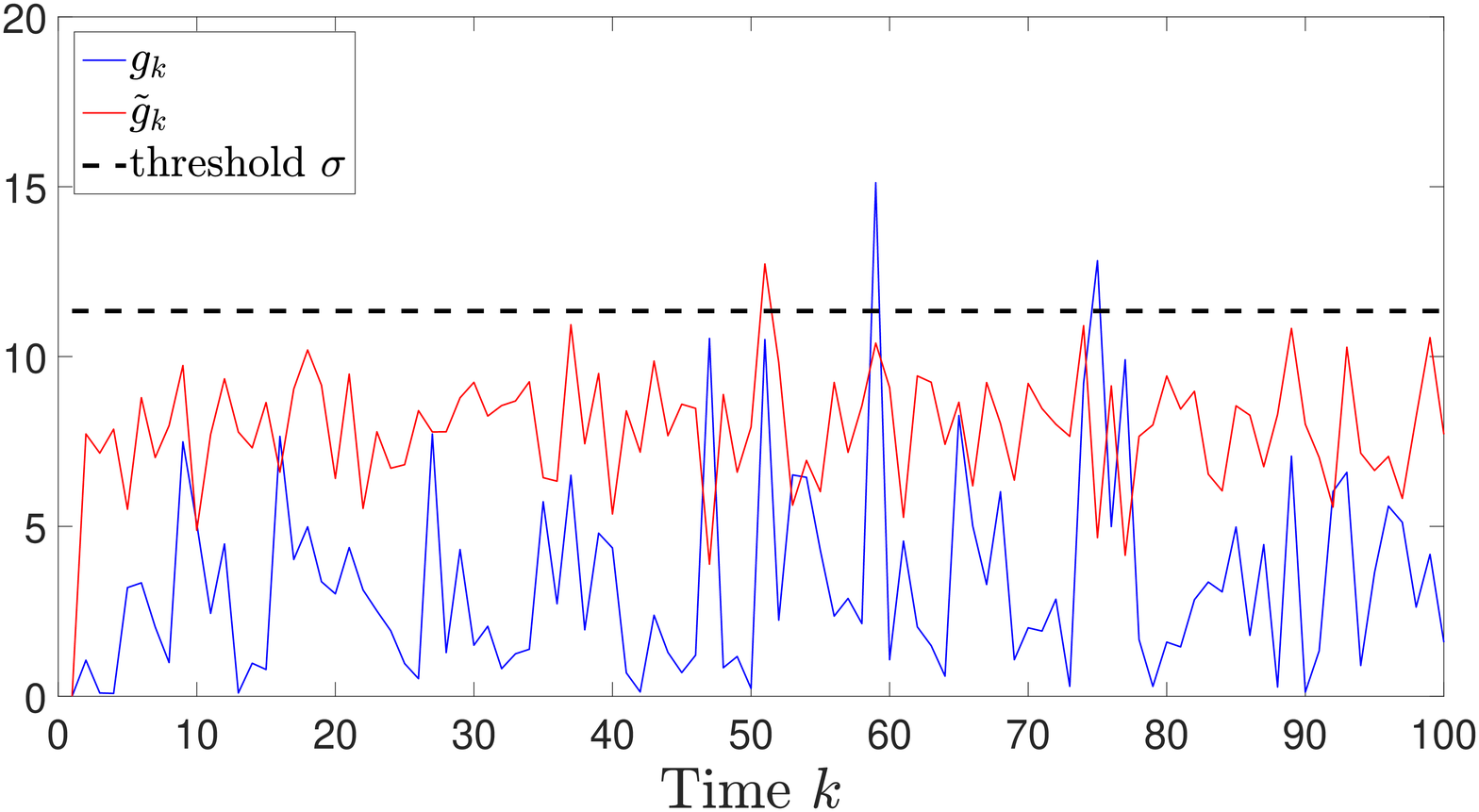}
	\caption{False data detection of $\chi^2$ detector.}
	\label{fig2}
\end{figure}

Fig. \ref{fig2} shows the alarming situation of the $\chi^2$ detector before and after attack. It can be noticed that the alarming rate is below $1\%$ in both two cases. In other words, the $\chi^2$ detector cannot distinguish this type of attack from the normal condition.

\begin{figure}[!t]
	\centering
	\includegraphics[width=3.5in]{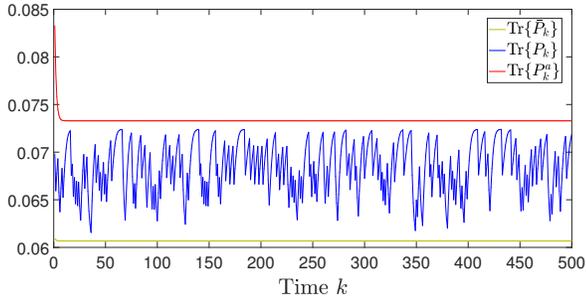}
	\caption{Remote estimation covariance for 3 cases.}
	\label{fig3}
\end{figure}

In Fig. \ref{fig3}, we examine the estimation error covariances under $3$ different cases, where $\bar{P}_k$, $P_k$, $P_k^a$ represent the covariances of standard Kalman filter, event-based estimator and estimation under attack, respectively. It can be noticed that not only suffering an estimation bias $\mathscr{E}$, the remote estimation under our attack strategy also has a larger error covariance.

\begin{figure}[!t]
	\centering
	\includegraphics[width=3.5in]{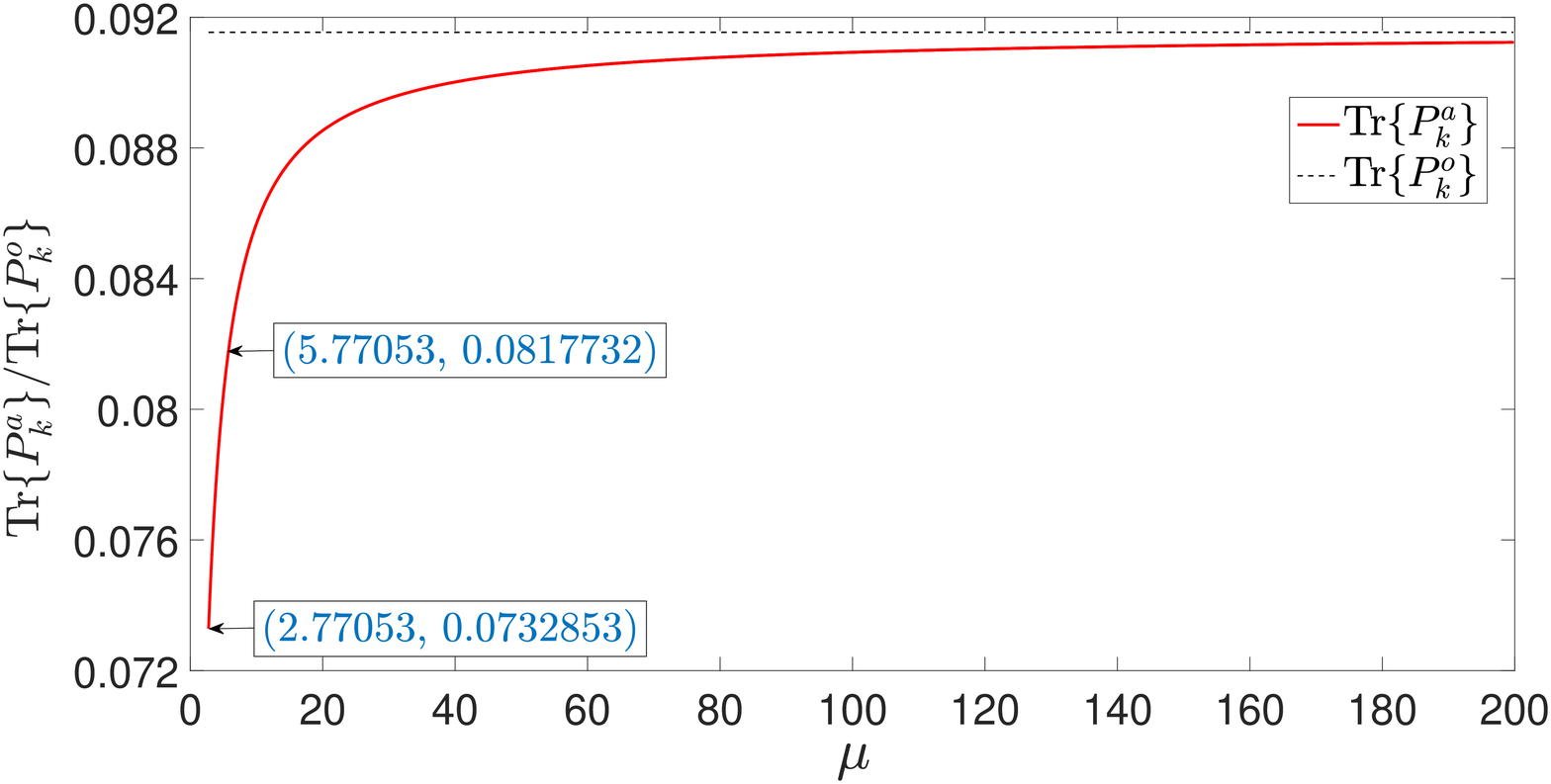}
	\caption{Steady state estimation covariance for different $\mu$.}
	\label{fig4}
\end{figure}

Fig. \ref{fig4} shows the variation tendency of the steady state estimation error covariance $P_k^a$ with the rise of scaling parameter $\mu$. We can find ${\rm Tr}\{P_k^a\}$ converges to $0.0915$ rapidly, which is the solution of (\ref{65}). In general, Fig. \ref{fig4} illustrates that system tends to be an open loop estimator if the attacker keep increasing $\mu$ and then proves the correctness of Corollary \ref{cor:3}.

\section{Conclusion}
In this paper, we proposed a novel deception attack strategy on event-based remote state estimator. The corresponding feasibility conditions were analyzed to ensure the attack can degrade the event-based scheduler performance while successfully bypassing the $\chi^2$ detector. Under the assumption that the event-based scheduler was triggered with probability $1$ after attack, we investigated the evolution of the remote estimation error covariance under the attack and analyzed the degradation of system performance. Furthermore, numerical simulations were presented to demonstrate the attack effectiveness.

\bibliographystyle{unsrt}        
\bibliography{root}           

\begin{thebibliography}{10}

\bibitem{poovendran2011special}
Radha Poovendran, Krishna Sampigethaya, Sandeep Kumar~S Gupta, Insup Lee,
  K~Venkatesh Prasad, David Corman, and James~L Paunicka.
\newblock Special issue on cyber-physical systems.
\newblock {\em Proceedings of the IEEE}, 100(1):1--12, 2012.

\bibitem{zhang2015optimal}
Heng Zhang, Peng Cheng, Ling Shi, and Jiming Chen.
\newblock Optimal denial-of-service attack scheduling with energy constraint.
\newblock {\em IEEE Transactions on Automatic Control}, 60(11):3023--3028,
  2015.

\bibitem{liu2011false}
Yao Liu, Peng Ning, and Michael~K Reiter.
\newblock False data injection attacks against state estimation in electric
  power grids.
\newblock {\em ACM Transactions on Information and System Security},
  14(1):1--33, 2011.

\bibitem{teixeira2015secure}
Andre Teixeira, Kin~Cheong Sou, Henrik Sandberg, and Karl~Henrik Johansson.
\newblock Secure control systems: A quantitative risk management approach.
\newblock {\em IEEE Control Systems Magazine}, 35(1):24--45, 2015.

\bibitem{mo2009secure}
Yilin Mo and Bruno Sinopoli.
\newblock Secure control against replay attacks.
\newblock In {\em Proceedings of the 47th annual Allerton conference on
  Communication, Control, and Computing}, pages 911--918, Monticello, IL, USA,
  October 2009.

\bibitem{mo2013detecting}
Yilin Mo, Rohan Chabukswar, and Bruno Sinopoli.
\newblock Detecting integrity attacks on scada systems.
\newblock {\em IEEE Transactions on Control Systems Technology},
  22(4):1396--1407, 2013.

\bibitem{guo2016optimal}
Ziyang Guo, Dawei Shi, Karl~Henrik Johansson, and Ling Shi.
\newblock Optimal linear cyber-attack on remote state estimation.
\newblock {\em IEEE Transactions on Control of Network Systems}, 4(1):4--13,
  2016.

\bibitem{pang2016two}
Zhong-Hua Pang, Guo-Ping Liu, Donghua Zhou, Fangyuan Hou, and Dehui Sun.
\newblock Two-channel false data injection attacks against output tracking
  control of networked systems.
\newblock {\em IEEE Transactions on Industrial Electronics}, 63(5):3242--3251,
  2016.

\bibitem{ni2019performance}
Yuqing Ni, Ziyang Guo, Yilin Mo, and Ling Shi.
\newblock On the performance analysis of reset attack in cyber-physical
  systems.
\newblock {\em IEEE Transactions on Automatic Control}, 65(1):419--425, 2019.

\bibitem{hu2018state}
Liang Hu, Zidong Wang, Qing-Long Han, and Xiaohui Liu.
\newblock State estimation under false data injection attacks: Security
  analysis and system protection.
\newblock {\em Automatica}, 87:176--183, 2018.

\bibitem{shi2014event}
Dawei Shi, Tongwen Chen, and Ling Shi.
\newblock Event-triggered maximum likelihood state estimation.
\newblock {\em Automatica}, 50(1):247--254, 2014.

\bibitem{wu2012event}
Junfeng Wu, Qing-Shan Jia, Karl~Henrik Johansson, and Ling Shi.
\newblock Event-based sensor data scheduling: Trade-off between communication
  rate and estimation quality.
\newblock {\em IEEE Transactions on Automatic Control}, 58(4):1041--1046, 2012.

\bibitem{nuttal1975some}
AH~Nuttall.
\newblock Some integrals involving the {$Q_M$} function.
\newblock {\em IEEE Transactions on Information Theory}, 21(1):95--96, 1975.

\bibitem{annamalai2009new}
Annamalai Annamalai, Chintha Tellambura, and John Matyjas.
\newblock A new twist on the generalized {Marcum $Q$-function $Q_M (a, b)$}
  with fractional-order ${M}$ and its applications.
\newblock In {\em Proceedings of the 6th IEEE Conference on Consumer
  Communications and Networking Conference}, pages 1--5, Las Vegas, NV, USA,
  February 2009.

\bibitem{sun2008inequalities}
Yin Sun and {\'A}rp{\'a}d Baricz.
\newblock Inequalities for the generalized {Marcum $Q$-function}.
\newblock {\em Applied Mathematics and Computation}, 203(1):134--141, 2008.

\bibitem{marcum1950table}
JI~Marcum.
\newblock Table of {$Q$}-functions, {U.S.} air force project {RAND} res. memo.
  {M-339, ASTIA} document {AD} 1165451.
\newblock Technical report, Rand Corp., Santa Monica, CA, USA, 1950.

\bibitem{sun2010monotonicity}
Yin Sun, {\'A}rp{\'a}d Baricz, and Shidong Zhou.
\newblock On the monotonicity, log-concavity, and tight bounds of the
  generalized marcum and nuttall {$Q$}-functions.
\newblock {\em IEEE Transactions on Information Theory}, 56(3):1166--1186,
  2010.

\bibitem{li2006computing}
Rong Li and Pooi~Yuen Kam.
\newblock Computing and bounding the generalized marcum ${Q}$-function via a
  geometric approach.
\newblock In {\em Proceedings of the IEEE International Symposium on
  Information Theory}, pages 1090--1094, Seattle, WA, USA, July 2006.

\bibitem{guo2018worst}
Ziyang Guo, Dawei Shi, Karl~Henrik Johansson, and Ling Shi.
\newblock Worst-case stealthy innovation-based linear attack on remote state
  estimation.
\newblock {\em Automatica}, 89:117--124, 2018.

\end{thebibliography}



\appendix
\end{document}